\documentclass[twocolumn]{article}
\let\OLDthebibliography\thebibliography
\renewcommand\thebibliography[1]{
  \OLDthebibliography{#1}
  \setlength{\parskip}{0pt}
  \setlength{\itemsep}{0pt plus 0.3ex}
}
\usepackage[dvipsnames]{xcolor}
\usepackage{hyperref}
\hypersetup{
    colorlinks=True,
    linkcolor=blue,
    filecolor=magenta,      
    urlcolor=purple,
    pdftitle={Overleaf Example},
    pdfpagemode=FullScreen,
    }
\usepackage{spconf,amsmath,graphicx}
\usepackage{amssymb}
\usepackage{comment}
\usepackage[utf8]{inputenc}
\usepackage{fourier} 
\usepackage{array}
\usepackage{makecell}
\usepackage{gensymb}
\usepackage{authblk}
\usepackage{subfigure}
\usepackage{lipsum}
\usepackage{caption}
\usepackage[english]{babel}
\usepackage[export]{adjustbox}
\DeclareMathOperator*{\argmax}{arg\,max}

\usepackage{graphicx} 
\usepackage{tabularx}
\usepackage{url}     
\usepackage{ragged2e}  

\title{Siamese Learning-based Monarch Butterfly Localization}

\name{%
\begin{tabular}{@{}c@{}}
Sara Shoouri$^{\star}$,
Mingyu Yang$^{\star}$,
Gordy Carichner$^{\star}$,
Yuyang Li$^{\dagger}$,
Ehab A. Hamed$^{\dagger}$,
Angela Deng$^{\star}$,
Delbert A. Green II$^{\star}$, 
Inhee Lee$^{\dagger}$,
David Blaaw$^{\star}$,
Hun-Seok Kim$^{\star}$
\end{tabular}}
\address{$^{\star}$ University of Michigan, Ann Arbor, MI, USA ,$^{\dagger}$ University of Pittsburgh, Pittsburgh, PA, USA}

\begin{document}
\maketitle
\begin{abstract}
A new GPS-less, daily localization method is proposed with deep learning sensor fusion that uses daylight intensity and temperature sensor data for Monarch butterfly tracking. Prior methods suffer from the location-independent day length during the equinox, resulting in high localization errors around that date. This work proposes a new Siamese learning-based localization model that improves the accuracy and reduces the bias of daily Monarch butterfly localization using light and temperature measurements. To train and test the proposed algorithm, we use $5658$ daily measurement records collected through a data measurement campaign involving 306 volunteers across the U.S., Canada, and Mexico from 2018 to 2020. This model achieves a mean absolute error of $1.416^\circ$ in latitude and $0.393^\circ$ in longitude coordinates outperforming the prior method.
\end{abstract}
\vspace{-0.25\baselineskip}
\vspace{-1\baselineskip}
\section{Introduction}
\label{sec:intro}
\vspace{-0.75\baselineskip}
GPS-less multi-sensor data geolocators \cite{bridge2013advances,lee2021msail} are miniaturized tracking devices that periodically collect sunlight, temperature, and other data to enable tracking of animals and even insects that migrate long ($>$1000 km) distances. Determining the daily location and trajectory of small animal and insect migrations is essential for understanding ecology, species interactions, and the impact of climate change on animals. For Monarch butterfly tracking, where a GPS receiver is not feasible because of its excessive power, size, and weight of the data logger, sunlight and temperature-based localization techniques have been investigated as a potential solution \cite{lee2021msail, yang2021migrating}.

Most prior work using light and temperature-based localization typically estimates the daily geo-coordinate based on a `threshold method' \cite{teo2004validation} or `template-fit' model \cite{philip2004advance}. In the threshold method, sunset and sunrise times are computed based on the moment when solar irradiance crosses a predefined threshold level \cite{lisovski2012geolocation}. Longitude is then estimated by the time of local noon and latitude by the measured day length. However, since latitude depends on the sun elevation angle, the estimation error is date- and location-dependent. The template-fit approach uses an analytical or data-driven template function for the location-dependent light variation to map the input data into latitude and longitude coordinates for a particular day. However, these methods typically have significant latitude ambiguity around equinox days due to the low day length variation everywhere on Earth.

To alleviate these issues, Yang \cite{yang2021migrating} presents a deep neural network (DNN)-based localization algorithm in which light and temperature data are fed into neural network models that estimate the daily position by approximating a likelihood probability. This algorithm can offer reduced estimation errors by combining the temperature and light-based probability estimation/heatmaps. However, it still has a relatively high latitude error, especially around the equinox when Monarch butterflies are actively migrating.

In this work, we propose a Siamese learning-based localization approach to estimate the location of the data loggers attached to Monarch butterflies collecting light intensity and temperature data. Our framework attempts to learn a general pattern representation of data and produce a pairwise similarity score for two given inputs, which quantifies the probability of proximity of their data collection locations. By evaluating the similarity between the sensor collected data (from unknown location) and the reference data (with known location) in the database, the proposed method estimates the daily butterfly location more reliably outperforming the prior art \cite{yang2021migrating}. 
We implement our algorithm for daily localization of Monarch butterflies that migrate from Canada to Mexico during September -- December each year. For training and testing the proposed network models, we use the data collected by a measurement campaign with $306$ volunteers \cite{Volunteer} to record the sunlight intensity and temperature from 2018 to 2020. Our method is applicable to the mSAIL \cite{lee2021msail} data logger, which is customized for Monarch butterfly attachment with a system size of $8 \times 8 \times 2.6$mm$^3$ and the weight of $62$mg (Fig. \ref{fig:Structure} top left). Our experimental results performed on the volunteer data show that the proposed algorithm can significantly reduce the localization error around the equinox (from $3.52^\circ$ to $1.74^\circ$) and increase the robustness of the results compared to the state-of-the-art \cite{yang2021migrating}. Our code is available at \url{https://github.com/sarashoouri/Siamese_Monarch}.

\vspace{-1\baselineskip}
\section{Method}
\label{sec:pipeline}
\vspace{-0.75\baselineskip}
\begin{figure*}[t]
\setlength\belowcaptionskip{-1.25\baselineskip}
 \centering
 \includegraphics[keepaspectratio,scale=0.18]{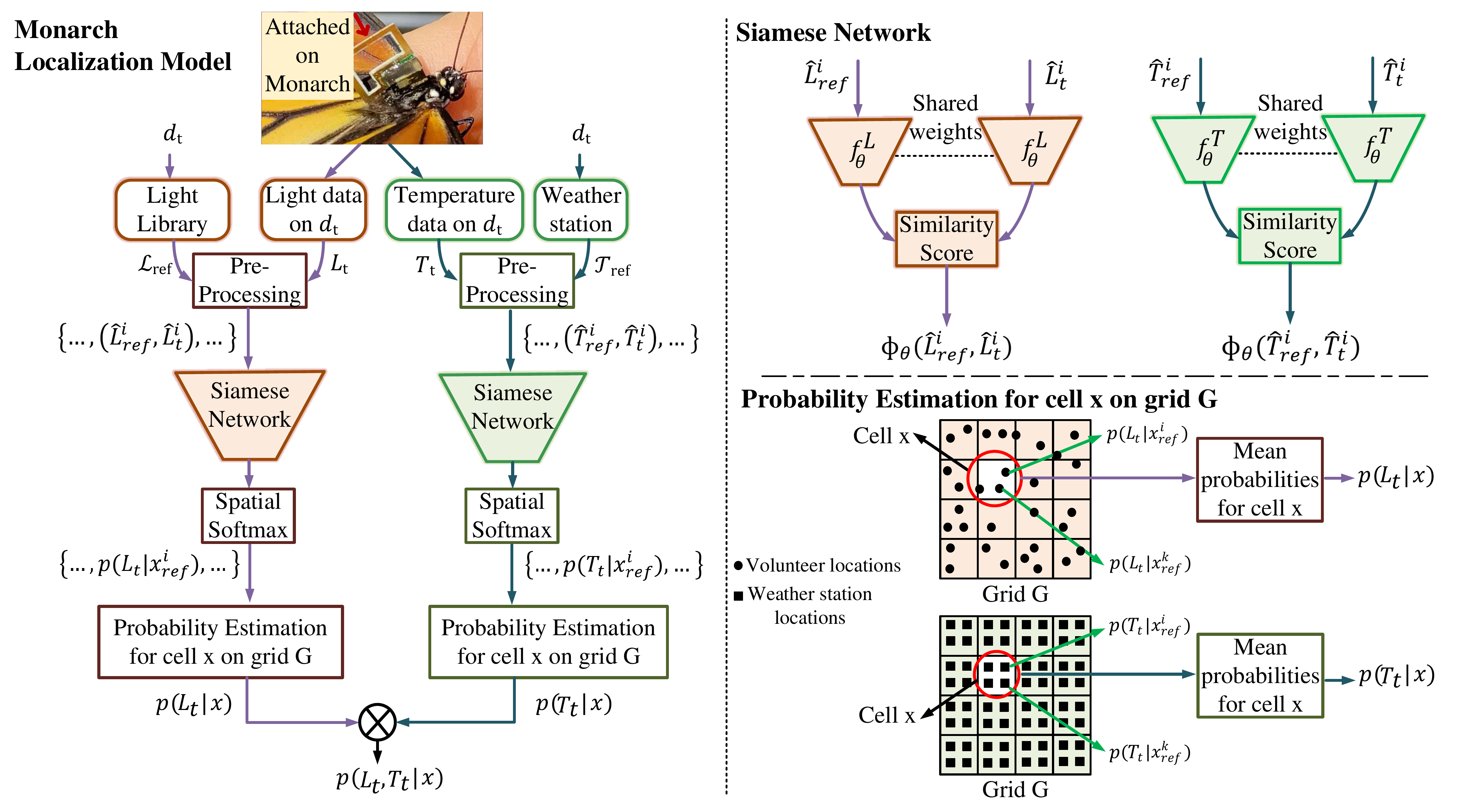}
 \vspace{-0.75\baselineskip}
 \caption{\textbf{Left:} Structure of the proposed algorithm. The sensor image is from \cite{lee2021msail}. \textbf{Top Right:} The Siamese network estimates the similarity between a pair of data. \textbf{Bottom Right:} $p(L_t|x)$ and $p(T_t|x)$ calculation for a given cell center $x$ on grid $G$.}
 \label{fig:Structure}
\end{figure*}

Our aim is to estimate the probability that a particular pair of light intensity $L_t$ and temperature $T_t$ measurements belongs to a $2D$ coordinate $x$ (latitude and longitude) on a specific date $d_t$. Thus, given a search grid $G$ containing all possible $2D$ coordinates, the estimated position $x_{est}$ on date $d_t$ can be computed using the maximum likelihood, $x_{est} = \argmax_{x\in G} \;  p(L_t, T_t|x)$. The likelihood value is estimated using two different Siamese neural network models: a light intensity model to compute $p(L_t|x)$, and a temperature model to compute $p(T_t|x)$. Similar to \cite{yang2021migrating}, 
we make a simplifying assumption that light and temperature generate conditionally independent probabilities for a given coordinate such that $p(L_t, T_t|x)\approx p(L_t|x)p(T_t|x)$ holds. We recognize that this assumption is not generally valid, but this simplification is necessary for our model training because of the data availability imbalance between the light and temperature data (i.e., it is difficult to fully characterize the correlation between them). While the temperature data is available from relatively densely populated weather stations, light intensity data is not measured/reported by weather stations. Hence, light intensity data is entirely obtained through volunteer data collections, which are only available for particular dates and on coarser coordinates.  

Our proposed algorithm has two main stages, 1) Data pre-processing, and 2) Contrastive Siamese learning \cite{bromley1993signature,koch2015siamese,lei2019siamese}-based localization. Fig. \ref{fig:Structure} shows the overall algorithm.
\vspace{-1.1\baselineskip}
\subsection{Data pre-processing}
\vspace{-0.65\baselineskip}
Monarchs rest during the night without moving. Thus, we perform daily localization utilizing the light and temperature data centered around ($\pm 9$ hours) the night center. The proposed data pre-processing method consists of two functions: a night center computing function $N$ and a time-shift function $r$. The night center computing function is designed separately for light ($N_L$) and temperature data ($N_T$). The function $r(I,N)$ time-shifts the data $I$ such that it is centered around the night center $N$.

To compute the night center for a given light intensity record $L$ for a day, we divide $L$ into two parts based on a hypothetical center point and calculate the cross-correlation between the two separated parts. Then, the hypothetical center point is adjusted until the cross-correlation reaches the maximum value, meaning that the divided parts have the highest symmetry. The time point that maximizes the cross-correlation value is assigned as the night center for $L$. Hence, the night center computing function $N_L$ takes $L$ as input and estimates the night center $n_c=N_L(L)$. 

Applying a similar method to the temperature measurement is unreliable since the temperature is often asymmetric around the night center. Hence, to align a temperature record $T$ of one day to its night center, we use an astronomical equation MATLAB function \cite{Equation} to calculate the night center $n_c$ for the location $x$ and date $d$. The night center $n_c$ for $T$ obtained at a (hypothetical) location $x$ and date $d$ is calculated by $n_c=N_T(x,d)$, where $N_T$ is the astronomical equation-based night center calculation function. 


We collect temperature data $T$ with a 1-hour interval to match the weather station's time resolution. We then produce a pre-processed temperature measurement $\hat{T}$ by time-shifting it (via $r$) to be centered around the night center: $\hat{T}=r(T,N_T(x,d))$ given a (hypothetical) location $x$ and a measurement date $d$. On the other hand, light intensity $L$ data for each day is collected from the sensor with a 1-minute interval, and is converted to log scale to emphasize the low light level variations around the sunrise and sunset. Note that the light intensity data collected from a sensor attached to a butterfly is prone to environmental variations that can lead to incorrect localization results. Thus, we first apply a denoising adversarial autoencoder (DAAE) \cite{makhzani2015adversarial} to $L$ to construct a denoised light record. The pre-processed light data $\hat{L}$, is thus obtained through the denoising DAAE ($\Psi$), night center ($N_L$), and time-shifting ($r$) calculation as follows: $\hat{L}=r(\Psi(L),N_L(\Psi(L)))$.

We now explain the denoising DAAE for the light intensity $L$. The goal is to estimate the clean data $\tilde{L}$ by an autoencoder $\Psi$ that consists of an encoder $\Psi_E$ and a decoder $\Psi_D$ pair that is trained to minimize a reconstruction loss. We treat this denoising as a distribution alignment task. The encoder $\Psi_E$ takes the noisy light $L$ to generate its latent representation $z$ with a smaller dimension, and the decoder $\Psi_D$ produces the denoised light $\tilde{L}$ from $z$. We desire to align the latent representations $z$ of noisy original data and $\tilde{z}$ of clean data by establishing a suitable discriminator $D$ to classify the latent vectors into original or denoised (clean) data. The autoencoder $\Psi(L) = \Psi_D(\Psi_E(L))$ and the discriminator $D$ are trained in an adversarial setting where each tries to outperform the other playing a two-player minimax game as initially introduced in \cite{goodfellow2014generative} using the loss:
\begin{equation}
\setlength\abovedisplayskip{4pt}
\setlength\belowdisplayskip{0pt}
    \mathcal{L}_{dis}=\mathbb{E}_{\tilde{z}}[log(D(\tilde{z}))] + \mathbb{E}_{z}[log(1-D(z))].
    \label{eq:loss_disc}
    \vspace{-1\baselineskip}
\end{equation}
\subsection{Siamese learning-based localization}
\vspace{-0.65\baselineskip}
\label{Siamese learning based localizations}
The prior published `template-fit' method \cite{teo2004validation} observed that the slope of light intensity has variations around sunrise and sunset that match at nearby locations on the same day. For example, the sun sets more slowly at high latitude vs. low latitude locations. We extend this idea to pattern matching and apply these slope variations to both light and temperature measurements. The primary assumption of our Siamese learning-based algorithm is that two closely located data on approximately the same day of the year (but in different years) should have highly correlated patterns. 

We first build a reference library of light intensities from the reference volunteer data with their known positions to perform the pattern matching. However, the constructed reference library is sparsely populated with only the locations where volunteers collected data, making it infeasible to perform pattern matching at arbitrary locations. We mitigate this issue by populating `synthesized' light data along with the longitude coordinate, significantly expanding the available light data for matching. The longitude coordinate mainly affects the time of the night center, and simply time-shifting the night center generates a synthesized light intensity at a new longitude coordinate. Thus, all of the volunteer data are time-shifted to cover our desired longitude range and construct the synthesized light intensity reference library. The amount of time-shift is 4 minutes per one degree of longitude. Note that we cannot apply a similar light intensity shifting to create synthesized data along the latitude coordinate as the light intensity pattern (after the night center alignment) per day depends on the latitude, and we cannot have a reliable model to `synthesize' it.

To describe the proposed Siamese learning-based localization, suppose $L_t$ is the light data collected on a known date $d_t$ at an unknown ground-truth 2D (latitude and longitude) location $x_t$. Let $\mathcal{L}_{ref}$ be a reference library (with volunteer collected and synthesized data). 
The reference light data subset $L_{ref}$ contains light data collected from $\pm 5$ days around $d_t$ at known 2D coordinates $x_{ref}$, and the size of this subset is denoted by $B$. Our goal is to quantify the similarity between each entry in $L_{ref}$ and $L_t$. Thus, for the $i$th element in $L_{ref}$, we obtained the pre-processed version $\hat{L}_{ref}^i=r(\Psi(L_{ref}^i),N_L(\Psi(L_{ref}^i)))$ after calculating its night center. Then, we generate a pre-processed version of the target light data $\hat{L}_{t}^i=r(\Psi(L_{t}),N_L(\Psi(L_{ref}^i)))$, using the same night center obtained from the reference $\Psi(L_{ref}^i)$. In this way, when the locations of the $L_{ref}^i$ and $L_{t}$ are matched, the estimated night centers align to each other, resulting in more accurate similarity scores. 


We then use a mapping function to convert the difference between $\{\hat{L}_{ref}^i, \hat{L}_{t}^i\}$ into a distance between their locations $\{x_{ref}^i, x_{t}\}$. For instance, if $\hat{L}_{ref}^i$ has a similar pattern as $\hat{L}_{t}^i$, then the distance between their locations $||x_{ref}^i - x_{t}^i||_2$ should be relatively small. 
We implement a Siamese neural network to construct this mapping function (transformer encoder) $\phi_{\theta}$, parameterized by $\theta$. Siamese networks are twin neural networks that share the identical weights \cite{bromley1993signature,koch2015siamese,lei2019siamese}. The function $\phi_{\theta}$ maps the pattern representation of the tuple $\{\hat{L}_{ref}^i, \hat{L}_{t}^i\}$ into a similarity score which indicates how close they are located. To construct $\phi_{\theta}$, we apply the same convolutional neural network $f_\theta^L$ to both $\hat{L}_{ref}^i$ and $\hat{L}_{t}^i$ to generate the feature vectors $f_\theta^L(\hat{L}_{t}^i)$ and $f_\theta^L(\hat{L}_{ref}^i)$ in the latent space. Euclidean distance between the latent vectors quantifies the similarity score between $\hat{L}_{ref}^i$ and $\hat{L}_{t}^i$ such that $\phi_{\theta}(\hat{L}_{ref}^i, \; \hat{L}_{t}^i)=||f_\theta ^L(\hat{L}_{ref}^i) - f_\theta^L(\hat{L}_{t}^i)||_2$.

To train the Siamese networks, positive (matching) and negative (non-matching) samples are required. Positive samples are the pairs of closely located data whose distance differences are less than $55km$ or $0.5^\circ$ in longitude and latitude, and the negative samples are the pairs of data whose distances are longer than $55km$. The model is trained using a contrastive loss function \cite{wu2017sampling} as described in Eq. (\ref{Contrastive}). The $y$ value is $1$ for the positive samples and $0$ for the negative samples, and $m$ is the threshold margin. The contrastive loss aims to maximize the similarity score for the positive samples while minimizing it for the negative samples.
\begin{equation}
\setlength\abovedisplayskip{6pt}
\mathcal{L}_{CL}=\frac{1-y}{2}||\hat{L}_{ref}^i - \hat{L}_{t}^i||^2_2+\frac{y}{2}\{max(0,m - ||\hat{L}_{ref}^i - \hat{L}_{t}^i||_2)^2\}.
\label{Contrastive}
\end{equation}
The spatial softmax function (\ref{eq:eq3}) is then applied to the similarity score between $\hat{L}_{ref}^i$ and $\hat{L}_{t}^i$ to convert it to a probability representation in the range of
$(0, 1)$. This represents the probability that $L_t$ is obtained at the location $x_{ref}^i$.
\begin{equation}
\setlength\abovedisplayskip{6pt}
    p(L_t |x_{ref}^i)=exp\left(-\frac{\phi ^2 (\hat{L}_{ref}^i, \; \hat{L}_{t}^i)}{2 \sigma ^2}\right) / \sum_{j=1}^{B} exp\left(-\frac{\phi ^2 (\hat{L}_{ref}^j, \;\hat{L}_{t}^j)}{2 \sigma ^2}\right).
    \label{eq:eq3}
\end{equation}
In (\ref{eq:eq3}), $\sigma$ is the standard deviation of the similarity scores for positive samples. The estimated probability (\ref{eq:eq3}) is evaluated with all $\{\hat{L}_{ref}^i, \hat{L}_{t}^i\}$ for $i=\{1,\cdots,B\}$ to create a set $I_{ref}$ containing the probabilities for given reference locations:
\begin{equation}
\setlength\abovedisplayskip{6pt}
  \begin{aligned}[b]
      & I_{ref}=\{(p(L_t |x_{ref}^1)
       ,x_{ref}^1),\cdots,
       &(p(L_t |x_{ref}^B), x_{ref}^B)\}.
  \end{aligned}
  \vspace{-0.5\baselineskip}
\end{equation}

After generating $I_{ref}$, the position $x_t$ of $L_t$ can be estimated by performing a coarse-to-fine grid search. A coarse search grid $G$ has $[27^\circ:1^\circ:48^\circ]$ latitude coordinate grids, and $[-122^\circ:1^\circ:-66^\circ]$ longitude coordinate grids. This search range covers our study area of the U.S, Canada, and Mexico. Each point on $G$ has a 2D coordinate $x$, and we compute $p(L_t|x)$ for all points on $G$. 

To estimate $p(L_t|x)$, we use the reference positions in $I_{ref}$ which are located around the position $x$. Thus, 
a subset of probabilities, called $I_{ref}^{cell}$, from $I_{ref}$ is created such that all points in $I_{ref}^{cell}$ satisfy $|x_{ref}-x|<1^\circ$. Finally, $p(L_t|x)$ is calculated by taking an average of the probabilities in the subset $I_{ref}^{cell}$. Although the reference library is densely populated along the longitude coordinate (due to the data `synthesis' procedure with time-shifting), the population along the latitude may be coarse for some regions where fewer volunteers were available. Thus, it is possible that with the constraint of $|x_{ref}-x|<1^\circ$, some grid cells centered at $x$ may have an empty $I_{ref}^{cell}$ set. For these empty cells, $p(L_t|x)$ is estimated based on the probabilities from nearest neighboring cells using linear interpolation. 

The spatial resolution of the
likelihood estimation for the grid $G$ is refined by upsampling and linear interpolating the heatmap of $p(L_t|x)$ with a $0.1^\circ$ resolution of $x$ on the refined $\tilde{G}$. The localization given $L_t$ (light-only localization) is completed through the maximum likelihood estimation on the refined $\tilde{G}$ such that $x_{t} \approx  \argmax_{\tilde{G}} \;  p(L_t|x)$.

The proposed method for the light data-based likelihood probability $p(L_t|x)$ estimation was extended to the temperature likelihood probability $p(T_t |x)$ estimation in a straightforward manner. Unlike the light data that is only available at sparse volunteer locations, the reference temperature data is available at all weather station locations which are densely distributed. We construct the temperature reference library $\mathcal{T}_{ref}$ by accessing the weather station data through WeatherBit API \cite{Weatherbit}. The pre-processed reference (weather station) data $\hat{T}_{ref}^i$ and sensor temperature data $\hat{T}_{t}^i$ on date $d_t$ are obtained by time-shifting both data using the night-center calculated by the reference temperature data, such that  $\hat{T}_{ref}^i=r(T_{ref}^i,N_T(d_t, x_{ref}^i))$ and $\hat{T}_{t}^i=r(T_{t},N_T(d_t, x_{ref}^i))$ hold. Note that denoising is not used for temperature data. A Siamese neural network $f_\theta^T$ is trained for temperature matching, then computes a similarity score between the pre-processed weather station temperature data $\hat{T}_{ref}^i \in \mathcal{T}_{ref}$ and the sensor data $\hat{T}_{t}^i$. This similarity score is then converted to a probability $p(T_t |x_{ref}^i)$ using Eq. (\ref{eq:eq3}) by replacing $L$ with $T$. The remaining steps to generate $p(T_t |x)$ are identical to the $p(L_t |x)$ estimation.

The final likelihood probability that jointly considers light and temperature data is approximated by the product: $p(L_t, T_t|x) \approx p(L_t|x)p(T_t|x)$ based on the simplifying assumption of conditional independence between the light and temperature measurements.
\vspace{-1.24\baselineskip}
\section{EXPERIMENTS}
\label{sec:EXPERIMENTS}
\vspace{-1.25\baselineskip}
\subsection{Data collection}
\label{ssec:Data Processing}
\vspace{-0.65\baselineskip}
We use real-world data collected through a data measurement campaign with $306$ volunteers across the U.S., Canada, and Mexico from 2018 to 2020 \cite{Volunteer}. Volunteers recorded light and temperature data using HOBO \cite{HOBO} sensors to emulate the mSAIL platform \cite{lee2021msail} (Fig. \ref{fig:Structure}, top left) from September to early December. The collected dataset contains $5658$ daily records with a time resolution of $10$ sec for light and $15$ sec for temperature. We use the year 2018 -- 2019 volunteer data as the training dataset (size of $3834$) and the year 2020 data as the testing set (size of $1824$). For temperature data, we use the weather station data with a time resolution of $1$ hour accessed through WeatherBit API \cite{Weatherbit}. Although the weather station data is much more densely populated than the volunteer data, it does not necessarily cover the entire study area. When the temperature data is not available for a particular location $x$, we apply the Kriging spatial interpolation \cite{zhang2017extended} to the nearby weather station data to obtain the temperature at that location.
\vspace{-1\baselineskip}
\subsection{Network structure}
\vspace{-0.65\baselineskip}
\begin{figure*}[t]
\setlength\belowcaptionskip{-1.5\baselineskip}
\centering
\begin{minipage}[t]{.32\textwidth} 
\centering
    \adjincludegraphics[valign=T,width=\textwidth]{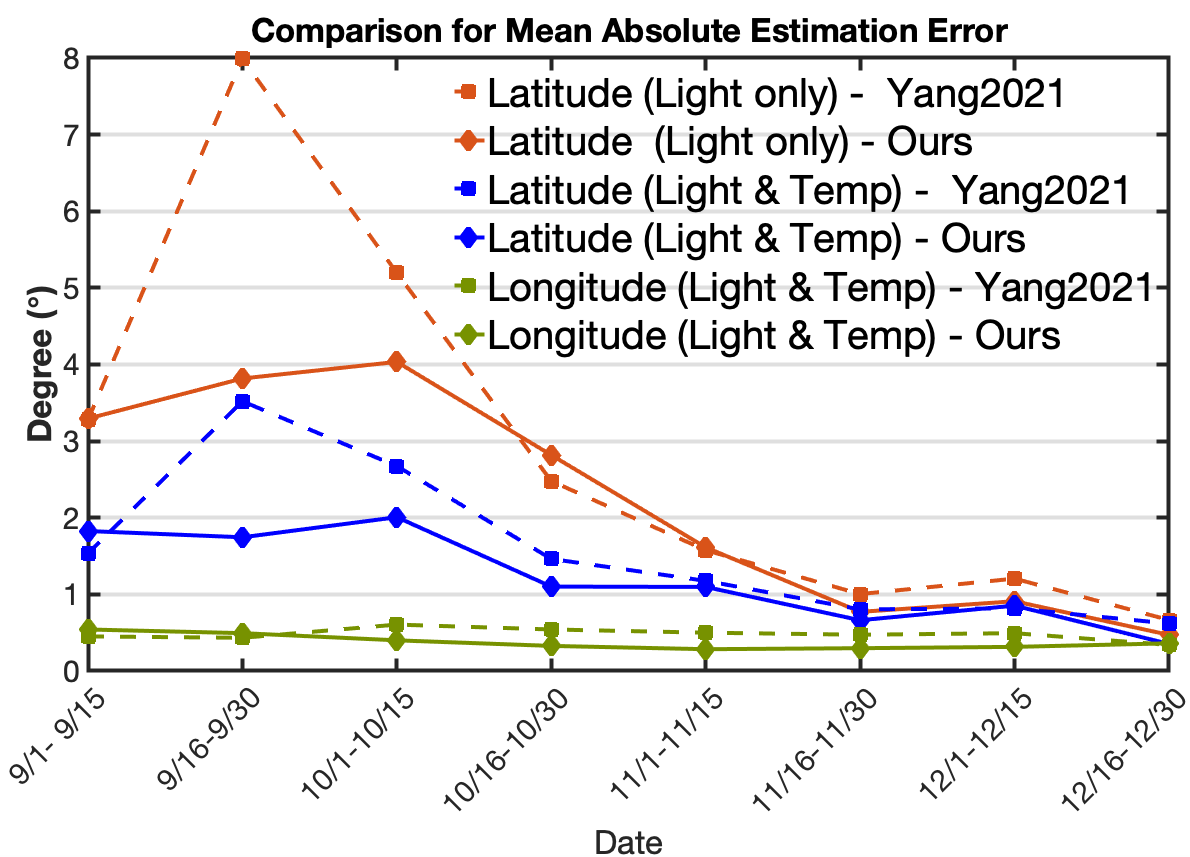}

\end{minipage}%
\quad 
\begin{minipage}[t]{.30\textwidth}
\centering
    \adjincludegraphics[valign=T,width=\textwidth]{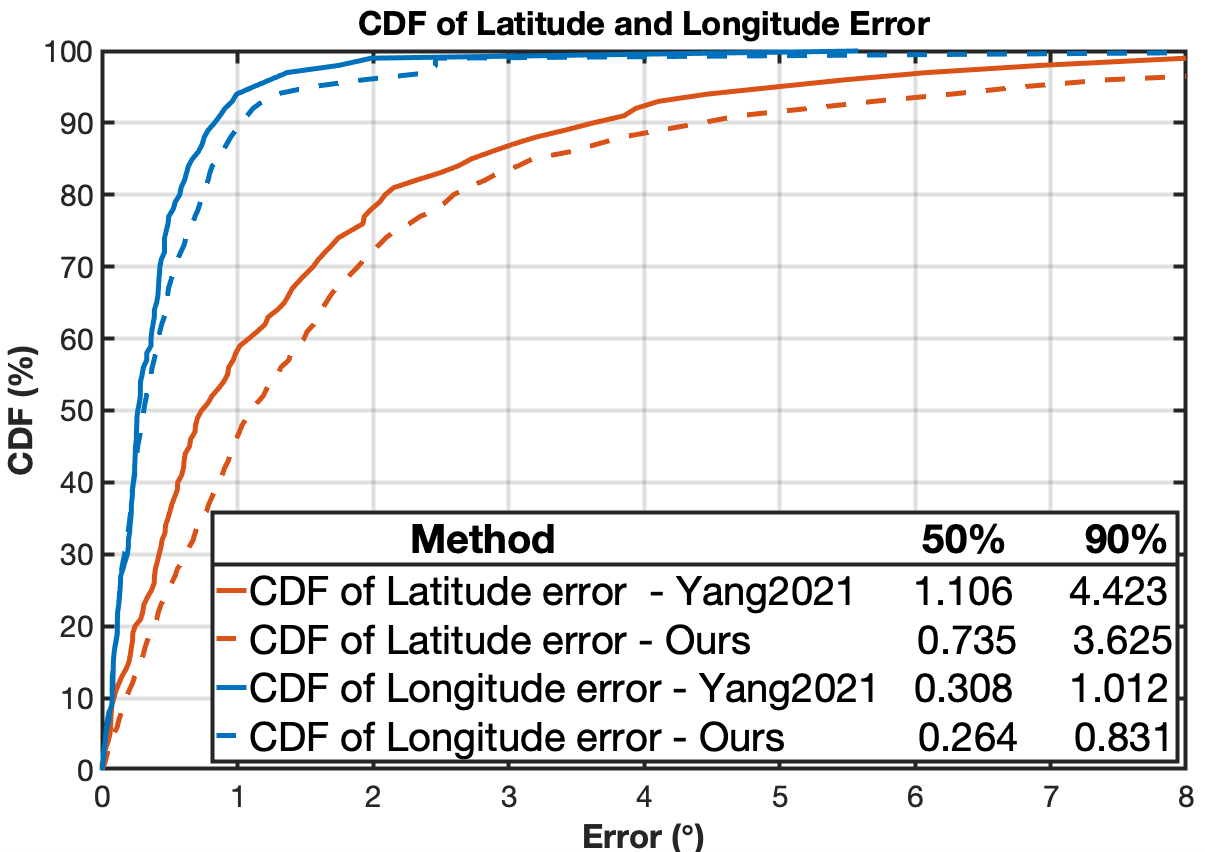}
\end{minipage}
\quad 
\begin{minipage}[t]{.31\textwidth}
\centering
    \adjincludegraphics[valign=T,width=\textwidth]{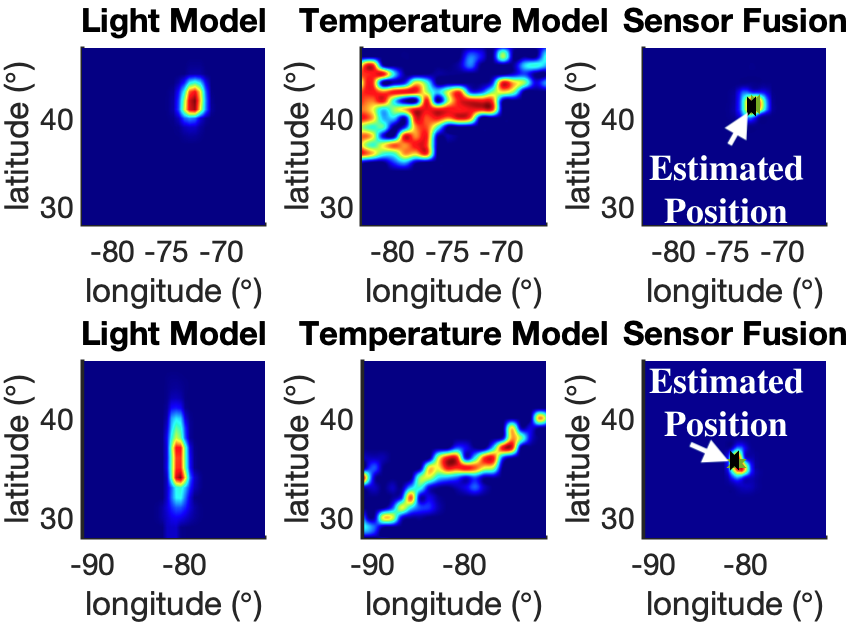}
\end{minipage}
\vspace{-0.75\baselineskip}
\caption{\textbf{Left}: MSE comparison evaluated biweekly for proposed method vs. Yang \cite{yang2021migrating}. \textbf{Middle}: Error CDF. \textbf{Right}: Example likelihood probability heatmaps for two different days. Red/yellow/blue indicates high/mid/low probability. First row: Dec 7, Second row: Nov 11. The ground-truth and estimated locations are marked with black and green points, respectively.}
\label{fig:test2}
\end{figure*}

The Siamese network $f_\theta^L$ for light data consists of $4$ convolutional (conv) layers containing convolution, batch normalization, ReLU, and max pooling, followed by $2$ fully connected layers (FCLs). A dropout layer of $p=0.30$ is applied after the first FCL. The size of the first conv layer is $128\times1\times9$ and the sizes of the other conv layers are $128\times1\times5$. The Siamese network for temperature data $f_\theta^T$ consists of $2$ conv layers with the size of $32\times1\times3$, and $2$ FCLs. A dropout layer of $p=0.30$ is used after the first FCL. The denoising encoder $\Psi_E$ (and decoder $\Psi_D$) consists of $3$ FCLs of size $480\times200$ ($50\times100$), $200\times100$ ($100\times200$), and $100\times50$ ($200\times480$), each followed by ReLU. The discriminator contains $3$ FCLs of size $50\times500$, $500\times500$, and $500\times1$, followed by Sigmoid. We use an ADAM optimizer with a learning rate of $0.001$ and a StepLR scheduler with a step size of $1000$ to train the models. At each epoch of the training for light data, we choose a batch size of two light records, whose date difference is less than $5$ days (but across different years). The temperature model is trained in a similar way, except that one input to the Siamese network is from the weather station. 
\vspace{-0.75\baselineskip}
\subsection{Localization Results}

\vspace{-0.65\baselineskip}
\label{sssec:Network structure}
We evaluate our test data localization on a $2D$ grid covering southern Canada, the U.S., and Mexico for Monarch butterfly localization. Fig. \ref{fig:test2} provides the
performance comparison between our proposed method and the state-of-the-art \cite{yang2021migrating} using the mean absolute error (MSE) evaluated biweekly. All results in this figure are based on the volunteer data (2018 - 2019 data for training and 2020 data for testing). The CDF of error in longitude and latitude degree is shown in Fig. \ref{fig:test2} middle. Our algorithm outperforms the baseline \cite{yang2021migrating}, especially around the equinox (Sep. 22), proving that the pattern matching technique can significantly compensate for the low day length variation issue around the equinox. Fig. \ref{fig:test2} shows that using the temperature data is also critical to compensate for the latitude estimation error from the light-only method around the equinox day when the night length is the same everywhere. Fig. \ref{fig:test2} (right subplot) visualizes the estimated probabilities $p(L_t|x)$, $p(T_t|x)$, and $p(L_t|x)p(T_t|x) \approx p(L_t, T_t|x)$.

\begin{figure}[!h]
\setlength\belowcaptionskip{-1\baselineskip}
    \centering
\vspace{-0.75\baselineskip}
   \includegraphics[keepaspectratio,width=0.65\linewidth]{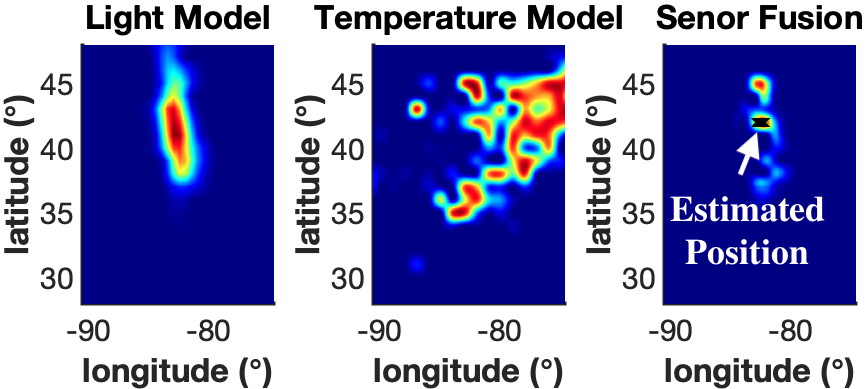}
\vspace{-0.75\baselineskip}
   \caption{Localization using butterfly-attached mSAIL \cite{lee2021msail} on Sep 17, 2021. Estimated point: $(42.0^\circ,-82.50^\circ)$. Ground-truth: $(42.032^\circ,-82.60^\circ)$}
\label{fig:Msail}

\end{figure}

We also evaluate our method using the data collected by an mSAIL \cite{lee2021msail} sensor attached to a wild Monarch butterfly on September 17, 2021 in Leamington, Ontario, near Lake Erie. The Monarch was released into the wild for a day, and then its sensor data was retrieved wirelessly while the butterfly was resting on a tree before flying across Lake Erie. Fig. \ref{fig:Msail} shows the localization result using the obtained data (the first hour of the data was extrapolated because the mSAIL sensor did not record it). The measured localization accuracy error is $0.032$ and $0.1$ degrees in latitude and longitude. 



\vspace{-1\baselineskip}
\subsection{Bias Evaluation}
\label{sssec:Bias evaluation}
\vspace{-0.65\baselineskip}

For a reliable unbiased localization, the error should not strongly depend on the number of available reference data near the ground-truth position. Learning such an unbiased method can be challenging as the training dataset is not uniformly populated and depends on the volunteer locations. To evaluate the robustness and bias of the localization model, we first quantify how densely the reference data is populated around a given ground-truth position by computing the average euclidean distance between the neighboring reference points and the ground-truth position. This is named the `Isolation score'. Then, we evaluate the correlation between the localization error and the computed Isolation score. Ideally, the error should be independent of the Isolation score, and the accuracy around isolated points should be as reliable as in more densely populated areas. We use three formulas, Pearson Correlation coefficient \cite{benesty2009pearson}, Distance Correlation \cite{szekely2007measuring}, and Mutual Information \cite{kraskov2004estimating} to measure the correlation between the error and Isolation score. Low numbers in these measures indicate that the method has less correlation and results have less bias towards low Isolation score locations. Table \ref{table:nBias Quantification Comparison} compares the measured correlation (or bias) between our model and the prior work \cite{yang2021migrating}. Our model has significantly less bias in all measures while producing lower errors (Fig. \ref{fig:test2}). Low correlation/bias scores from our model imply that the localization error depends less on the density of reference (training/testing) data around the ground-truth location. 
\vspace{-0.75\baselineskip}

\begin{table}[h!]

\caption{Measured bias of Yang 2021 \cite{yang2021migrating} and our model.} 
\vspace{-0.75\baselineskip}
\centering
\large
\resizebox{0.45\textwidth}{!}{
\begin{tabular}{c c c c} 
\hline\hline 
\thead{\Large{\textbf{Method}}} & \thead{\large{\textbf{Pearson}} \\ \large{\textbf{Correlation}} \\\large{(latitude, longitude)}} & \thead{\large{ \textbf{Distance}} \\ \large{\textbf{Correlation}}\\\large{(latitude, longitude)}} & \thead{\large{\textbf{Mutual}}\\ \large{\textbf{Information}} \\\large{(latitude, longitude)} }\\ [0.5ex] 
\hline 
\\
\textbf{\Large{Yang2021 \cite{yang2021migrating}}} & (0.26, 0.324) & (0.23, 0.292) & (0.33, 0.397) \\ 
\thead{\Large{\textbf{Our model}}} & \textbf{(0.070, 0.172)}
 & \textbf{(0.0744, 0.167)}
 & \textbf{(0.245, 0.308)}\\
\hline
\end{tabular}}
\label{table:nBias Quantification Comparison} 
\vspace{-1\baselineskip}

\end{table}
\vspace{-1\baselineskip}
\section{Conclusion}
\vspace{-0.75\baselineskip}
We developed a Siamese learning-based localization using light intensity and temperature measurements for miniaturized data loggers to study Monarch butterfly migration. The proposed method significantly outperforms the prior method, especially around the equinox. Moreover, it has less bias towards more densely populated areas and achieves lower accuracy errors. The proposed algorithm demonstrates the successful localization of a wild butterfly using real-world data. The presented model exhibits a mean absolute error of $1.416^\circ$ in latitude and $0.393^\circ$ in longitude for the volunteer collected test dataset.
\vspace{-1\baselineskip}
\subsubsection*{ACKNOWLEDGMENTS}
\vspace{-0.5\baselineskip}
This work was in part funded by NSF IIBR Award $\#$2045017 and National Geographic Society Grant. We thank all the volunteers who participated in the data measurement campaign for this research.

\vspace{-0.5\baselineskip}

\bibliographystyle{rand}
\bibliography{refs}
\newpage
\appendix
\setcounter{secnumdepth}{-1}
\section{Supplemental Material}
\begin{figure}[!h]
\setlength\belowcaptionskip{-1\baselineskip}
    \centering

   \includegraphics[keepaspectratio,width=0.9\linewidth]{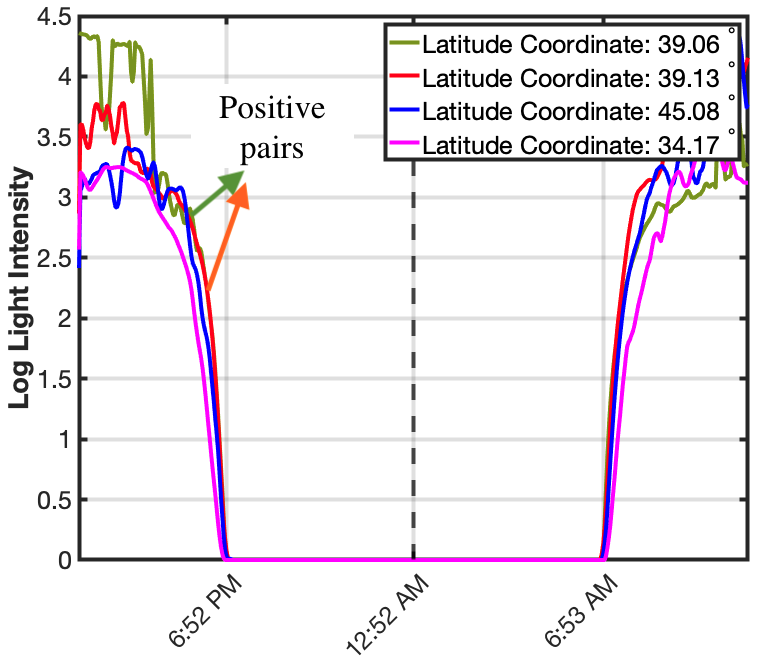}
\vspace{-0.75\baselineskip}
   \caption{Light intensities at various latitude coordinates with the same longitude coordinate.}
\label{fig:Appendix}

\end{figure}

Fig. \ref{fig:Appendix} displays the slope of light intensity variations around the sunset and sunrise around the equinox day. The light measurements are located at the same longitude coordinate with different latitude coordinates. The night length (time duration when the log of light intensity is below 0) is equal regardless of the latitude. It can be observed that the light records with close latitude coordinates (red and green light data) have matching slopes and patterns before the sunset and after the sunrise, whereas the slopes are different when latitude coordinates are mismatched. Our algorithm exploits this similarity/difference to improve the localization performance near equinox days.

\end{document}